\documentstyle[11pt]{article}
\begin{document}

\begin{flushright}
Liverpool Preprint: LTH 348\\
 hep-lat/9504016\\
24th April, 1995\\
\end{flushright}

%\vspace{5mm}
 
\begin{center}
{\LARGE\bf
Lattice sum rules for the colour fields
}\\[5mm]

{\bf  C. Michael  }\\

{DAMTP, University of Liverpool, Liverpool, L69 3BX, U.K.}\\
% cmi @ liv.ac.uk
\end{center}

\begin{abstract}

We analyse the sum rules describing the action and 
energy in the colour fields around glueballs, torelons and 
static potentials. 

\end{abstract}

\section{Introduction}

In lattice gauge theory, it is possible to derive sum rules for the 
energy and action in the colour fields around states.  The technique
used in ref\cite{mich} --- hereafter  referred to as I --- involves
evaluating derivatives with respect to a parameter ($\beta$ for example)
of the formal expression for a  correlation involving an observable of
intetest. For  the Wilson gauge action, this yields exact relations
between the  $\beta$-dependence of observables and the sum over a
time-slice of the plaquette expectation  in the presence of that
observable (for example see eqs.~\ref{sr1} and \ref{sr2}).   These
identities can be used  as checks of numerical  results --- see
ref\cite{gl} for  an application to the gluelump state.  They relate the
variation with $\beta$ to sums at one  fixed value of $\beta$.  They can
also be used to investigate the $\beta$-dependence  of lattice
quantities: so leading to evaluation of the lattice beta-function.

A more powerful set of relations can be derived  if the
$\beta$-derivatives can be re-expressed using renormalisation  group
invariance in terms of well known quantities.  For glueballs  and
potentials, these were also presented in I. The main conclusion is  that
the combination of squared colour field strengths corresponding  to the
action (electric plus magnetic) is much larger than the  combination
corresponding to the energy (electric minus magnetic).  This implies
that the electric and magnetic field strengths are  comparable. This
conclusion has been a useful benchmark for models  of  non-perturbative
QCD.  Although this general conclusion was correct, the explicit results
in I were in error and the correct  expressions are given here.

The lattice analysis of the field strengths depends on the scale  at
which these fields are probed. The results can be calculated  reliably
by perturbation theory for very short distance scales. However, for
scales appropriate for non-perturbative states,  perturbation theory on
the lattice in terms  of the bare coupling is now known to be poorly
convergent and it is  worthwhile to re-assess the assumptions leading to
these relations.

Recently there has been a reawakening of interest in this area - partly
from new accurate lattice results~\cite{bali} and partly because of the
realisation~\cite{dosch,rothe} that the application~\cite{mich} of the
sum rules  to static potentials was wrong.

Here we summarise the derivation of I and confirm the correction
needed for the application to  potentials. We extend the  discussion to
apply the sum rules to torelons and to analyse  the transverse and
longitudinal colour fields separately.   This enables us to explore in
detail the problem of the field energy in  the potential between static
sources.

\section{Glueballs}

As an example of the techniques to be used, we consider first a glueball
state. Define $M(\beta)$ as the lattice observable glueball mass (in
lattice units) which will depend on  $\beta$ the bare lattice  coupling
parameter, with $\beta=2N/g^2$, for the gauge sector of  the SU(N)
theory.

Then, for the Wilson action, the identity was derived in I that
\begin{equation}
   {d M \over d \beta}
=<1\,|\sum \Box |\,1> - <0\,|\sum \Box |\,0>
 = \sum \Box_{1-0}
  \label{sr1}
\end{equation}
 where $\Box$ is the plaquette action ${1 \over N}{\rm Tr}(1-U_{\Box})$
which  is summed over all ($6L^3$) plaquettes in one time slice. The 
subscript $1-0$ refers to the difference of this plaquette sum in  a one
glueball state (1) and in the vacuum (0).  

This identity can be used as it stands to check this observed plaquette
difference with the left hand side obtained as a finite difference  from
lattice calculations of $M(\beta)$ at two nearby values of $\beta$. A
more  powerful application comes from using the renormalisation group
invariance  to relate the $\beta$-dependence of $M$ to the $\beta$-
dependence of  the lattice spacing $a$. Since $M(\beta(a))/a$ is the 
physical continuum mass $m$ as $a \to 0$,  it must be independent of
$a$. Hence
 \begin{equation}
0 = {dM(\beta(a))/a \over da} = -{M \over a^2} + 
{ 1 \over a}{ dM \over d \beta}\, {d\beta \over da}
\end{equation}
Thus 
\begin{equation}
M=  {d\beta \over d \ln a} \,  \sum \Box_{1-0}
\label{MA}
\end{equation}
Note that $d\beta / d\ln a= -11 N^2/(12 \pi^2)$ to lowest order 
in perturbation theory for an SU(N) gauge theory. Thus the plaquette 
action is {\it lowered} in the glueball surroundings compared to the 
vacuum.

This is one of the prototype lattice action sum rules. It relates  the
plaquette action around a glueball to the mass of the  glueball. It 
is exact provided that a non-perturbative determination of the lattice 
beta function is used.

Further relations can be derived by splitting the lattice Wilson  action
into several terms with different coefficients. This analysis  of
asymmetric lattices was incorrect in I. In order to  establish clearly
the correct expressions, here we use a more direct  method of derivation
which also has the advantage of being more general.

  Consider the  general case where there are different coefficients for
all 6  orientations of plaquette:
 \begin{equation}
\beta \sum_{i,j,i<j} \Box_{ij} \to \sum_{i,j,i<j} \beta_{ij} \Box_{ij}
 \end{equation}
There will be four lattice spacings $a_i$ in general. We shall need 
to evaluate the derivatives $\partial \beta_{ij} / \partial a_k$. At 
the symmetry point where $a_i=a$ for all $i=1,4$,  these 
derivatives fall into two classes 
 \begin{equation}
{\partial \beta_{ij} \over \partial \ln a_k}=
S \ \hbox{if}\ k=i \ \hbox{or}\ j
\ \ \hbox{and} \ \
{\partial \beta_{ij} \over \partial \ln a_k}=
U \ \hbox{if}\ k\ne i \ \hbox{or}\ j
\end{equation}

The generalisation of the identities derived in I are also needed:
\begin{equation}
   {\partial M \over \partial \beta_{ij}} = \sum (\Box_{ij})_{1-0}
\end{equation}
 where the sum is again over one time slice.
 Then the renormalisation group invariance of the result obtained 
on such a lattice implies that
\begin{equation}
{\partial \over \partial a_i} {M(\beta_{jk}(a_0,a_1,a_2,a_3),..)
 \over a_0} = 0
\label{rge}
\end{equation}
 where $a_0$ enters because the glueball correlation is conventionally 
determined in the time direction. Because only this time direction is
privileged  in this case, at the symmmetry point, we have that
 \begin{equation}
\Box_{0j}=\Box_t \ \ \hbox{and} 
\ \ \Box_{jk}=\Box_s \ \ \hbox{for}\ j,k \ne 0
\end{equation}
where the subscript 1-0 is implied hereon.

Applying the renormalisation group invariance conditions 
of eq.~\ref{rge} for $i=0$ and for $i \ne 0$ gives
\begin{equation}
M= \sum 3 S \Box_t + 3 U \Box_s 
\label{MASU}
\end{equation} 
\begin{equation}
0= \sum (2U+S)  \Box_t + ( U+2S) \Box_s 
\label{M0SU}
\end{equation}
 Then combining eq.~\ref{MASU} and eq.~\ref{M0SU} yields
\begin{equation}
M= \sum 2 (S +U) (3\Box_t + 3  \Box_s) 
\label{MGSU}
\end{equation} 
which is the the same as eq.~\ref{MA} provided
we have the consistency condition 
 \begin{equation}
2(S+U) =  {d\beta \over d \ln a} 
\label{BETA}
\end{equation}
Subtracting eqs.\ref{MASU} and \ref{M0SU} then gives
\begin{equation}
M= \sum {2 \over 3} (S-U) (3\Box_t - 3\Box_s) 
\label{MESU}
\end{equation} 

This latter equation is appropriate to the energy in the colour 
field around a glueball. In order to make it more useful, we 
need to estimate the combination of derivatives $S-U$.

Consider the special case, as used by Karsch~\cite{karsch}, where
$a_t=a_0;\ a_1=a_2=a_3=a_s$ and $\beta_{0i}=\beta_t;\ \beta_{ij}=
\beta_s$ where $i,j>0$.  The derivatives in this case can be related to
$S$ and $U$, at the symmetry point:
 \begin{equation}
{\partial \beta_t \over \partial  \ln a_t}=S
\ \hbox{and} \
{\partial \beta_t \over \partial  \ln a_s}= S+2U
\label{BT}
\end{equation}
 \begin{equation}
{\partial \beta_s \over \partial  \ln a_t}=U
\ \hbox{and} \
{\partial \beta_s \over \partial  \ln a_s}= 2S+U
\label{BS}
\end{equation}

 The  dependence of $\beta_s$ and $\beta_t$ 
on $a_t$ and $a_s$ coming from the weak coupling limit of the 
theory~\cite{karsch} is that, where $\xi=a_s/a_t$, 
 \begin{equation} \beta_t=\xi (\beta(a_s) + 2Nc_s(\xi) +.. ) \
\hbox{and} \ \beta_s= \xi^{-1}(\beta(a_s) + 2Nc_t(\xi) +.. )
\label{KAR}
\end{equation} 
 where at $\xi=1$: $c_s=c_t=0$ and Karsch obtains $c_s'=0.114$ for $N=2$
and $c_s'=0.2016$ for  $N=3$. Then substituting eqs.~\ref{KAR} into
eqs.~\ref{BT} and \ref{BS} gives the constraint
 \begin{equation}
4N(c_t' + c_s')=-{d \beta \over d \ln a}
\end{equation}
This is the same constraint as found by Karsch from similar consistency 
arguments.

Using these expressions gives
 \begin{equation}
S=-\beta+ 2Nc_s'+ {1 \over 2} {d \beta \over d \ln a}
\ \hbox{,}\ \ \ 
U=\beta- 2Nc_s'
\end{equation}
and thus
\begin{equation}
  S-U=
-2\beta+ 4Nc_s'+ {1 \over 2} {d \beta \over d \ln a}
\end{equation}
 This implies that, as $\beta \to \infty$,  the energy sum-rule
(eq.~\ref{MESU}) becomes
 \begin{equation}
M= \sum {4 \over 3} \beta (3\Box_s - 3\Box_t) 
\label{MESR}
\end{equation}

The expression of eq.~\ref{MESR} in I had a factor of 1 instead of 4/3,
coming from an error in the evaluation of the weak coupling result 
for the dependence of  the asymmetric $\beta$'s on the $a$'s. Note that
the naive continuum expression for the energy in the colour field would
be obtained with a factor of 1. 

In principle $S-U$ can be determined non-perturbatively by simulating  a
lattice with non-equal $\beta$'s and determining the ratio of the
lattice  spacings in the 4 directions from the glueball correlations  in
those directions. Accurate data do not exist at present, although an
indirect method has been used in SU(2) and substantial corrections are
found~\cite{ekr} to the  weak coupling results.   This is not surprising
since lattice perturbation theory in the bare coupling is now  known to
be poorly convergent. This non-perturbative  evaluation~\cite{ekr} gives
values of $(U-S)/(2\beta)$ of 0.66 at  $\beta=2.4$ and 0.77 at
$\beta=2.8$ compared to the weak coupling values  of 0.85 and 0.87 
respectively.   It is amusing that these non-perturbative estimates  are
close to 0.75 which would give the naive energy relation: 
 \begin{equation} 
M= \sum  \beta (3\Box_s- 3\Box_t)  
\label{MESRT} 
\end{equation}

The gluonic vacuum in QCD is known to be polarisable. It behaves  like a
medium and  can be assigned an effective  dielectric constant. Thus it
is not really surprising that the  naive sum of  the energy in the
colour fields (ie $\Sigma  \beta (3\Box_s- 3\Box_t)$ ) does not agree
exactly with the mass. Indeed the result will  depend in QCD on the
scale at which the field energy is evaluated.  A sensible scale would be
commensurate with the glueball mass - where  a non-perturbative
determination of $S-U$ is needed and rough agreement is  obtained
between the apparent field energy and the mass. The weak  coupling
calculation (which shows that only  3/4 of the mass lies  as apparent
energy in the colour fields) implies a very  short distance scale of
energy determination - which will probe  the vacuum polarisation in a
different manner.

It is worth emphasizing the basic result, which was obtained in I
already, that the electric and magnetic field strengths are comparable.
In detail, the departure from equality is correctly given by
 \begin{equation}
{ {\cal E} \over{\cal B} } = 
{\Box_t \over \Box_s}
=  {- U-2S \over 2 U+S} 
\approx 1-{3 \over 2\beta} {d  \beta \over d \ln a }
\approx 1+ {33 N \over 12 \pi}\, {g^2 \over 4 \pi}
\approx 1
\end{equation}
 where the approximation used in estimating $S$ and $U$ is valid at
large $\beta$.

\section{Potentials and torelons}

Having calibrated the approach on the glueball, we consider string 
states. The potential between static quarks is the case of greatest 
practical interest. Another related situation is with a closed loop of 
colour flux encircling the periodic boundary condtions: the torelon.
As discussed in I, there is some subtlety in  principle in dealing with
the self energy of the static quarks.   For clarity of presentation, 
the torelon case is considered first since the derivation is more compact.

%Thus a convenient  step is to consider a torelon state first.

The torelon is a closed string of colour flux in the fundamental 
representation that encircles the periodic boundary conditions  in the
$x$-direction where there are $R$ lattice spacings  in this direction.
Its energy is measured on a lattice by analysing  correlations of 
closed Polyakov line operators at $t=0$ and $t=T$. The study of the
large  $T$-behaviour then gives the lattice observable $E(R,\beta)$.

The analysis of I gives, where $R$ is kept constant,
\begin{equation}
 \left. {\partial E(R) \over \partial \beta} \right| _R = \sum \Box_{1-0}
  \label{sr2}
\end{equation}
 where `1' now refers to the plaquette expectation value between  torelon
states and the sum is again over all plaquettes in a time-slice. The
renormalisation group analysis  now needs to take  account of the  fact
that $r=Ra$ must be kept constant in taking the limit $a \to 0$. So
 \begin{equation}
0 = \left. {dE(R,\beta(a))/a \over da}\right|_r = 
-{E \over a^2} -{R \over a^2}{\partial E \over  \partial R}
+ { 1 \over a} {d\beta \over da}\,\left.
{\partial  E \over \partial \beta}\right|_R
\end{equation}
Thus 
\begin{equation}
 E(R)+{\partial E \over  \partial \ln R}=  {d\beta \over d \ln a} \, 
\sum \Box_{1-0}
 \label{TA}
\end{equation}
 As pointed out by Dosch et al.~\cite{dosch}, this expression differs
from that in I where the term with the derivative with respect to $R$
was omitted.  The net effect of that term is, for a confining potential,
to increase the effective left hand side of the action sum rule by a
factor of 2. 

We now apply the general consideration of 6 couplings $\beta_{ij}$ as
above.  The new feature is that the torelon correlator is extended in
the $x$ and $t$ directions.  Thus we need to distinguish the $x$
(longitudinal: L) and $y,\ z$ (transverse: P) spatial directions.  The 4
independent types of plaquette have orientations $tL$, $tP$, $LP$, and
$PP$: we label them as ${\cal E}_L$, ${\cal E}_P$, ${\cal B}_P$, ${\cal
B}_L$ respectively in a natural notation.  Note that ${\cal E}$ here is
related to the difference of the plaquette value in the torelon state
and in the vacuum and so is the difference of gauge invariant
combinations of electric colour fields squared.  Following the same
steps as above we obtain three independent constraints from the
invariance with respect to $a_0$, $a_L$ and $a_P$ of ${1 \over a_0}
E(Ra_L, \beta_{ij}(a_k))$
 \begin{equation} E= \sum  S {\cal E}_L + 2 S {\cal
E}_P  + 2 U {\cal B}_P +U {\cal B}_L  
\label{T0SU} \end{equation} 
 \begin{equation} R {\partial E \over \partial R}=  \sum  S {\cal E}_L +
2 U {\cal E}_P  + 2 S {\cal B}_P +U {\cal B}_L  
\label{T1SU}
\end{equation}  
 \begin{equation} 0= \sum  U {\cal E}_L + (S+U) {\cal
E}_P  + (S+ U) {\cal B}_P +S {\cal B}_L  
\label{T2SU} 
\end{equation} 
 where the sum is over one time slice.

Combining these equations we obtain
\begin{equation}
E+R {\partial E \over \partial R}= 
\sum 2(S+U) ( {\cal E}_L + 2  {\cal E}_P  + 2  {\cal B}_P + {\cal B}_L) 
\label{TASU}
\end{equation} 
\begin{equation}
E+R {\partial E \over \partial R}= 
\sum 2(S-U) ( {\cal E}_L - {\cal B}_L) 
\label{TELSU}
\end{equation} 
\begin{equation}
E-R {\partial E \over \partial R}= 
\sum 2(S-U) (  {\cal E}_P  -   {\cal B}_P ) 
\label{TEPSU}
\end{equation} 
 It is also convenient to write down the combination corresponding 
naively to the total energy in the fields:
 \begin{equation} E- {1 \over 3} {\partial E \over \partial \ln R}= 
\sum {2 \over 3}(S-U) ( {\cal E}_L + 2  {\cal E}_P  - 2  {\cal B}_P -
{\cal B}_L)  
\label{TESU} 
\end{equation} 
 Again the action sum-rule (eq.~\ref{TASU}) agrees with the result 
obtained from a symmetric lattice (eq.~\ref{TA}) with the same 
relationship of $S+U$ to the beta function (eq.~\ref{BETA}) as 
for the glueball case.
 Thus, apart from the term with a derivative with respect to $R$, the
results  for the total action (eq.~\ref{TASU}) and total energy
(eq.~\ref{TESU}) are similar in normalisation to the  glueball case
introduced above.

Consider, for orientation, the case where the torelon energy is 
a sum of a string tension piece and a string fluctuation piece:
\begin{equation}
E(R)= K R - f/R
\label{ER}
\end{equation} 
 where we expect $f=\pi/3$ (this behaviour of the torelon energy has
been checked numerically recently~\cite{ms}).  Then the sum rules become
 \begin{equation}
2KR=
\sum 2(S+U) ( {\cal E}_L + 2  {\cal E}_P  + 2  {\cal B}_P + {\cal B}_L) 
\label{KASU}
\end{equation} 
\begin{equation}
2KR=
\sum 2(S-U) ( {\cal E}_L - {\cal B}_L) 
\label{KELSU}
\end{equation} 
\begin{equation}
-2f/R=
\sum 2(S-U) (  {\cal E}_P  -   {\cal B}_P ) 
\label{KEPSU}
\end{equation} 
 This shows that the transverse energy in the fields (where here we 
define energy  as  ${\cal E} - {\cal B}$ ) will be much smaller for large
$R$ than the longitudinal energy. Moreover it  has the opposite sign.
These sum rules provide an independent way to  study the split of the
total torelon energy into string tension and  string fluctuation
components.

Consider the sum rule for the longitudinal energy with the weak coupling 
($\beta \to \infty$) value for $S-U$:
\begin{equation}
{1 \over 2}KR=
\sum \beta ( {\cal B}_L - {\cal E}_L) 
\label{KEL}
\end{equation}
 The right hand side is just the naive expression for the energy in the
longitudinal fields.  Thus we obtain one half of the expected
semi-classical result of $KR$.  This is somewhat surprising since the
longitudinal colour flux is applied explicitly and in the semi-classical
limit the energy should remain relatively unaffected by quantum
corrections.  However, the vacuum polarisation effects will be strong at
the large energy scale (corresponding to $\beta \to \infty$) used to
evaluate the expression. 

The application to the potential between static sources follows the 
same steps as for the torelon. The difference is that $R$ is now the 
spatial extent of a Wilson loop rather than the spatial extent of the 
lattice itself. The main new feature is that there will also be a 
self-enegry contribution in the lattice observable energy $E(R)$.  This 
self-energy was discussed in I. As $a \to 0$ it becomes the dominant 
term in the energy but it is very localised spatially.  Thus it is
possible to separate it out  --- leaving just the  same
results as for the torelon discussion above. One way to  remove the
self energy contribution, in practice, is by taking the  difference of
expressions for two values of $R$ when it cancels.

The analogue of eq.~\ref{ER}  for  the potential energy $E(R)$ between 
static sources at separation $R$ is a sum of 
self-energy, Coulombic and string tension terms:
 \begin{equation}
  E(R)=V_0 -e/R +KR
 \end{equation}
 After removing the self-energy part ($V_0$), the sum rules then become 
the same as eqs.~\ref{KASU}, \ref{KELSU} and \ref{KEPSU} with $f$ 
changed to $e$.  Thus the same result applies that the transverse 
energy in the colour fields will be much smaller for large $R$ than  the
longitudinal energy (where here we  define energy  as  ${\cal E} -
{\cal B}$ ). This is a new result.

\section{Conclusions}

We have studied the energy and action distribution in the colour fields
around non-perturbative states.  We use the semi-classical definition of
these distributions and define the appropriate difference of the
plaquette combination evaluated in the non-perturbative state and in the
vacuum.  The lattice definition we use has energy and action determined
by  the plaquette. In principle it should be possible to define a
quantity  which characterises the energy and action in the colour fields
of a state  and which has a continuum limit. Such a defintion could be
based, for example, on using  a square  Wilson loop of {\it fixed}
physical size as $a \to 0$. This  would probe the energy and action
distributions  at a {\it fixed} physical scale and so would  give a 
result free of lattice artefacts. But, of course, such a definition 
would not satisfy  the sum rules we have derived.

The simplest result, which was obtained in I and which has been
checked in numerical studies, is that the action in the colour fields is
much larger than the energy.  This follows because the derivatives of
$\beta_{ij}$ with respect to $a_k$ on an asymmetric lattice can be
expressed in terms of two independent quantities $S$ and $U$ which can
be estimated.  The correct expression for the ratio, derived here, is
 $$ {{\rm Energy} \over {\rm Action}} = 
{ 3(S+U) \over (S-U)} 
\approx  {-3 \over 4\beta}\, {d \beta \over  d \ln a}
\approx {g^2 \over 4 \pi}\, {33N \over 24 \pi}
 << 1
 $$
 for a glueball state (where the approximation used in estimating $S$
and $U$ is valid at large $\beta$). For the interquark potential or for
a torelon this ratio is approximately 3  times smaller still.

The naive expectation is that the spatial sum of the energy density in 
the colour field around a state should equal the energy of the state
itself. We correct the results given in I, and   find that, evaluated by
the semi-classical expression, the sum of the energy density in the
colour fields around a glueball (and torelon) is given by 3/4 (and 1/2
respectively) of the energy of the state times $2.0 \beta/(U-S)$.  Thus
no non-perturbative value for $S-U$ can make both of these sums exactly
equal to the energy.  Evaluating the energy density sum at a large
energy scale, we can use  perturbation theory to obtain a field energy
around a glueball (and torelon) which is 3/4 (and 1/2 respectively) of
the energy of the state.  These fractions are closer to one when a
non-perturbative  estimate at a lower energy scale is used to determine
the field sums.  The explanation for the departure of these relations
from identity is most easily achieved by invoking the vacuum
polarisation effects as producing an effective dielectric constant. 
Moreover this effective dielectric constant must be different in the
glueball (spherical) and torelon (cylindrical) geometries. 

We also present sum rules for the longitudinal and transverse  field
energy in a string state: torelon or interquark potential.  For the 
potential between static quarks, this implies that at large $R$ the
transverse energy in the colour fields will  be much smaller than the
longitudinal energy. It will be interesting to explore this in lattice
studies.

\end{document}